\documentclass[acmlarge]{acmart}
\usepackage{xspace} 
\usepackage{pifont}
\usepackage{hyperref}
\usepackage{enumitem}
\usepackage{subcaption}
\usepackage{lscape}
\usepackage[procnames]{listings}
\usepackage{color}
\usepackage{tabularx}
\usepackage{multirow}
\definecolor{keywords}{RGB}{255,0,90}
\definecolor{comments}{RGB}{0,0,255}
\definecolor{red}{RGB}{255,0,0}
\definecolor{green}{RGB}{0,150,0}

\AtBeginDocument{%
  \providecommand\BibTeX{{%
    \normalfont B\kern-0.5em{\scshape i\kern-0.25em b}\kern-0.8em\TeX}}}

\begin{document}


\title{Can ChatGPT Write a Good Boolean Query for Systematic Review Literature Search?}


\author{Shuai Wang}
\affiliation{%
	\institution{The University of Queensland}
	\city{Brisbane}
	\country{Australia}}
\email{shuai.wang2@uq.edu.au}

\author{Harrisen Scells}
\affiliation{%
  \institution{Leipzig University}
  \city{Leipzig}
  \country{Germany}}
\email{harry.scells@uni-leipzig.de}

\author{Guido Zuccon}
\affiliation{%
	\institution{The University of Queensland}
	\city{Brisbane}
	\country{Australia}}
\email{g.zuccon@uq.edu.au}

\author{Bevan Koopman}
\affiliation{%
	\institution{CSIRO}
	\city{Brisbane}
	\country{Australia}}
\email{bevan.koopman@csiro.au}

\newcommand\todo[1]{{\color{red}#1}}
\newcommand{\cgpt}{ChatGPT\xspace}
\newcommand{\metha}{unguided\xspace}
\newcommand{\methb}{guided\xspace}
\newcommand{\tick}{\ding{51}}
\newcommand{\cross}{\ding{55}}

\begin{abstract}

Systematic reviews are comprehensive reviews of the literature for a highly focused research question. These reviews are often treated as the highest form of evidence in evidence-based medicine, and are the key strategy to answer research questions in the medical field. To create a high-quality systematic review, complex Boolean queries are often constructed to retrieve studies for the review topic. However, it often takes a long time for systematic review researchers to construct a high quality systematic review Boolean query, and often the resulting queries are far from effective. Poor queries may lead to biased or invalid reviews, because they missed to retrieve key evidence, or to extensive increase in review costs, because they retrieved too many irrelevant studies. Recent advances in Transformer-based generative models have shown great potential to effectively follow instructions from users and generate answers based on the instructions being made. In this paper, we investigate the effectiveness of the latest of such models, ChatGPT, in generating effective Boolean queries for systematic review literature search. Through a number of extensive experiments on standard test collections for the task, we find that ChatGPT is capable of generating queries that lead to high search precision, although trading-off this for recall. Overall, our study demonstrates the potential of ChatGPT in generating effective Boolean queries for systematic review literature search. The ability of ChatGPT to follow complex instructions and generate queries with high precision makes it a valuable tool for researchers conducting systematic reviews, particularly for rapid reviews where time is a constraint and often trading-off higher precision for lower recall is acceptable.

\end{abstract}
\maketitle
\keywords{Boolean query, Systematic Review, Prompt Engineering, ChatGPT }


\section{Introduction}
To construct a high-quality systematic review, all evidence related to a research topic is examined by the researchers of the review; those relevant to the research topic will be selected and further evaluated and synthesised.
To gather all evidence related to the review, Boolean queries are authored to search medical databases. Boolean queries provide reproducibility, explainability, and the benefit of filtering out articles not relevant to the research topic~\cite{macfarlane2022search}, therefore reducing the workload of unnecessary document assessments.
However, constructing a high-quality Boolean query is challenging, even for experienced searchers, and the resulting queries are often sub-optimal~\cite{scells2018generating}. 

To help researchers construct better systematic review Boolean queries, automatic Boolean query formulation and refinement methods have been developed~ \cite{scells2018generating,scells2019refinement,scells2020automatic,scells2020comparison,scells2020computational,wang2022automated,wang2021mesh,badami2022systematic,scells2020sampling,pourreza2022towards, wang2022automated, wang2022mesh}. With formulation, we refer to the task of creating a Boolean query from scratch, at times relying on one or more ``seed'' documents; i.e., example studies related to the systematic review. With refinement, we refer to the task of improving an existing Boolean query; e.g., by adding or removing Boolean clauses and operators to yield a higher precision (less irrelevant documents retrieved) while maintaining recall. Although results have been promising and methods have been integrated into tools for assisting in the query creation process~\cite{scells2018searchrefiner,li2020systematic,scells2022impact,wang2022mesh}, automated techniques are yet far from being able to yield high-quality Boolean queries.


Recent advances in text generation models~\cite{li2022survey,gozalo2023chatgpt} have led to great successes in task-based question-answering and provide high-quality responses based on user intentions~\cite{li2022survey}. One of such generative models, ChatGPT\footnote{\url{ https://openai.com/blog/chatgpt/}}, is considered the most expressive text-generation model at the time of writing, and achieved state-of-the-art effectiveness across natural language processing tasks.
This paper investigates how to use ChatGPT to construct high-quality systematic review Boolean queries effectively. The following research questions guide our investigation into using ChatGPT:

\begin{itemize}
	\item[\bf RQ1:] How does ChatGPT compare with current state-of-the-art methods for formulating and refining systematic review Boolean queries?
	\item[\bf RQ2:] To what extent do the prompts used to generate systematic review Boolean queries impact the effectiveness of the Boolean queries produced by ChatGPT?
	\item[\bf RQ3:] What is the effect of guiding the query formulation process with ChatGPT through multiple prompts that mimic the process of the current state-of-the-art automated Boolean query generation methods?
	\item[\bf RQ4:] What are the caveats and potential challenges of using ChatGPT to create systematic review Boolean queries? 
\end{itemize}

To our knowledge, this is the first such attempt to comprehensively evaluate the effectiveness of using ChatGPT for creating Boolean queries. This includes carefully engineering prompts for ChatGPT and thoroughly evaluating using standardised test collections for systematic reviews. 

We find that ChatGPT compares favourably with current state-of-the-art query generation methods. Improvements in precision sometimes come at the expense of a recall, although we show that this may be mitigated with better MeSH term handling and the practice of snowballing (locating more studies via citation networks). We show generating a good prompt is paramount. Providing a sample Boolean query as part of the prompt is beneficial. Multiple prompting interactions with ChatGPT are better than a single interaction. While results are promising, they are also volatile in that effectiveness varies across runs and prompts; MeSH suggestion, in particular, is poor with ChatGPT.

The findings of this research can help systematic review researchers use ChatGPT for systematic review Boolean query construction and refinement, and understand its limitations and caveats.

\section{Related Work}

\subsection{Systematic Review Automation}

Constructing systematic reviews requires manual effort by trained professionals, across several distinct phases, with different types of manual effort required at each phase. 

Phase 1 is to define a specific research question; software can help in defining a good querstion~\cite{cochrane2019revman}. 

Phase 2 is to author Boolean queries based on the research question. The query defines which (and how many) results are returned, so their quality greatly impacts all later phases. This is why we focus specifically on query generation. Automated methods have been developed for query formulation~\cite{russell20182dsearch,clark2020improving,li2020systematic}, query refinement~\cite{scells2018searchrefiner,li2020systematic} and query exploitation~\cite{demner2007answering,boudin2010clinical}. 

Phase 3 is to screen all the titles and abstracts retrieved by the query, manually assessing them for relevance. The primary automation approach to help here is screen prioritisation~\cite{anagnostou2017combining,norman122017limsi,hollmann2017ranking,singh2017iiit,yu2017data,minas2018aristotle,norman2018limsi,di2018interactive,di2017interactive,lee2017study,singh2017identifying,scells2017ltr,cohen2018uic,alharbi2017ranking,alharbi2018retrieving,chen2017ecnu,wu2018ecnu,martinez2008facilitating,karimi2009challenge}, which involves ranking the studies according to their likely relevance to the research question. Once ranked, automated approaches can define a `stopping criteria', after which screening does not continue because all relevant documents are likely to be found. Active learning approaches~\cite{cohen2006reducing,miwa2014reducing} are often employed during screening to help rank include studies to be screened. 

Phase 4 involves extracting specific details from studies relevant to the review; data extraction methods can help here~\cite{summerscales2009identifying,hsu2012automated}. 

Phase 5 is the synthesis of all the evidence into a single coherent review document. Synthesis automation~\cite{summerscales2011automatic,rodriguez2009figure,thomas2008methods,animesh2018neuralparscit,clark2020full} have been developed here; they include text thematic analysis~\cite{torres2017revmanhal} and even text generation~\cite{rathbone2017automating}.

\subsection{Automatic Query Formulation and Refinement}

\textbf{Query formulation} is the process of deriving a Boolean query, based on the research question, according to a specific set of guidelines. The guideline presents two accepted procedures that exist for developing a query for systematic reviews. These procedures describe the steps one should take when developing a query. The first procedure is called the \textit{conceptual} method~\cite{suhail2013methods}. This procedure requires first identifying several high-level concepts from pilot searches or potentially relevant studies known a priori. These high-level concepts are then used to discover synonyms and related keywords. The query is then iteratively refined, a process driven by the expertise of the information specialist. The second procedure is called the \textit{objective} method~\cite{simon2010identifying,hausner2015development}. The first step of this procedure is to create a small set of potentially relevant studies that will seed the rest of the procedure, e.g., through pilot searchers, as in the conceptual method. Multiple statistical procedures follow, which extract terms from these studies and weakly validate the query. The identified terms must still be added to the query manually, which, like the conceptual method, is driven by the expertise of the information specialist. The exact details of these methods are unimportant for understanding their use in this paper; they should be understood simply as a series of steps one can follow to arrive at a query.

Naturally, these methods require a considerable amount of time (as the information specialist must perform multiple pilot searches and spend time validating the search)~\cite{karimi2009challenge,reeves2002twelve} and are prone to human error~\cite{salvador-olivan2019errors,sampson2006errors}. To this end, Scells et al.~\cite{scells2020comparison} investigated automating these two query formulation procedures. The main finding of this line of research was that computationalising these procedures could not match the effectiveness of humans; however, further manual refinement of the automatically generated queries dramatically improved retrieval effectiveness. 

\textbf{Automatic Query Refinement} was developed based on the observed benefit from manual query refinement. Such methods take an initial human authors query and apply a series of transformations (adding terms or clauses) to make the query more effective~\cite{scells2018searchrefiner,li2020systematic}. In combination with query visualisation~\cite{scells2018searchrefiner} tools, these query refinement tools were able to improve the initial query.

Learnings from the query formulation approach drive two clear directions for using \cgpt to automate query formulation: the first is to allow \cgpt to generate queries however it sees fit, and the second is to guide \cgpt by prompting it to follow the instructions of the conceptual or objective procedures. We refer to the first method as \textit{\metha} and the second method as \textit{\methb}. Learnings from the query refinement work made us hypothesise that providing an existing query to \cgpt and asking for a refinement could be beneficial.

\subsection{Prompt Engineering for Transformer-based Generative Language Models}

Prompt engineering is the process of guiding a generative language model to perform a particular task. In some respect, prompt engineering can be seen as a way to program a generative language model through natural language~\cite{reynolds2021prompt}. A popular way of guiding model output through prompt engineering is for text-to-image generative models~\cite{liu2022design,oppenlaender2022taxonomy}.
This `zero-shot' or `few-shot' approach to tasks with generative language models has also achieved state-of-the-art results on several natural language tasks~\cite{brown2020language}. More recently, prompt engineering has been applied to natural language tasks for medicine, such as question answering~\cite{luo2022biogpt,lievin2023can}.
The use of generative language models for query formulation or refinement is relatively under-explored. We are only aware of a single other paper that has published results about using a generative language model for query expansion~\cite{claveau2021neural}, and no work has examined this method for creating Boolean queries for systematic literature review search.

\begin{table*}[t!]
	\centering
	\small
	\begin{tabular}{p{3pt}cp{380pt}}
		\toprule
		&Prompt ID & Prompt \\
		\midrule
		
		\multirow{3}{*}{\rotatebox{90}{Simple}}
		&q1& For a systematic review titled “\{review\_title\}”, can you generate a systematic review Boolean query to find all included studies on PubMed for the review topic? \\
		&&\\ \midrule
		
		\multirow{11}{*}{\rotatebox{90}{Detailed}}
		&q2& You are an information specialist who develops Boolean queries for systematic reviews. You have extensive experience developing highly effective queries for searching the medical literature. Your specialty is developing queries that retrieve as few irrelevant documents as possible and retrieve all relevant documents for your information need. Now you have your information need to conduct research on \{review\_title\}. Please construct a highly effective systematic review Boolean query that can best serve your information need. \\ \cmidrule{2-3}
		
		&q3&Imagine you are an expert systematic review information specialist; now you are given a systematic review research topic, with the topic title “\{review\_title\}”. Your task is to generate a highly effective systematic review Boolean query to search on PubMed (refer to the professionally made ones); the query needs to be as inclusive as possible so that it can retrieve all the relevant studies that can be included in the research topic; on the other hand, the query needs to retrieve fewer irrelevant studies so that researchers can spend less time judging the retrieved documents. \\ \midrule
		
		\multirow{11}{*}{\rotatebox{90}{With Example}}
		&q4& You are an information specialist who develops Boolean queries for systematic reviews. You have extensive experience developing highly effective queries for searching the medical literature. Your specialty is developing queries that retrieve as few irrelevant documents as possible and retrieve all relevant documents for your information need. You are able to take an information need such as: “\{example\_review\_title\}” and generate valid pubmed queries such as: “\{example\_review\_query\}".  Now you have your information need to conduct research on “\{review\_title\}”, please generate a highly effective systematic review Boolean query for the information need. \\ \cmidrule{2-3}
		
		&q5&You are an information specialist who develops Boolean queries for systematic reviews. You have extensive experience developing highly effective queries for searching the medical literature. Your specialty is developing queries that retrieve as few irrelevant documents as possible and retrieve all relevant documents for your information need. A professional information specialist will extract PICO elements from information needs in a common practice in constructing a systematic review Boolean query. PICO means Patient/ Problem, Intervention, Comparison and Outcome. PICO is a format for developing a good clinical research question prior to starting one’s research. It is a mnemonic used to describe the four elements of a sound clinical foreground question. You are able to take an information need such as: “\{example\_review\_title\}" and you generate valid pubmed queries such as: “\{example\_review\_query\}". Now you have your information need to conduct research on “\{review\_title\}”. First, extract PICO elements from the information needs and construct a highly effective systematic review Boolean query that can best serve your information need. \\
		\bottomrule
	\end{tabular}
	
	\caption{Prompts for single prompt query formulation}
	\vspace{-23pt}
	\label{table:query_formulation_prompt}
\end{table*}

\section{Engineering Prompts for Systematic Reviews}
In this paper we investigate the use of ChatGPT to create queries for systematic review literature search. The basic mechanism employed by ChatGPT is that to take an input sequence of text (called \textit{prompt}), process it through the model, and output the next token in the sequence. This process is repeated several times to generate a complete response to the prompt. ChatGPT relies upon the Transformer architecture~\cite{vaswani2017attention} trained on a massive amount of text data, allowing it to learn patterns in the way that language is used. During the generation process, the model uses these learned patterns to generate text that is similar to the text observed at training.

A key component in the use of and interaction with ChatGPT is the design of the prompt used to instruct the model to generate an answer. To instruct ChatGPT to create Boolean queries for systematic review literature search we iteratively devise a number of prompts of increased complexity, including prompts that rely on example Boolean queries. We also experiment with prompts that guide the generation of the query through multiple interactions (guided prompts), and by leveraging insights from the existing objective method of query formulation~\cite{hausner2012routine,scells2020comparison,scells2020computational}. Prompts are devised for two tasks related to Boolean query generation: the formulation task and the refinement task. 


\vspace{-10pt}
\subsection{Single Prompts for Query Formulation}
\label{single_p_query_generation}
Single query formulation prompt refers to a prompt that instructs to formulate systematic review Boolean queries using the title of the review. 
We design five such prompts for the Boolean query formulation task; these prompts are reported in Table \ref{table:query_formulation_prompt}. They can be classified into three categories: \textit{simple}, \textit{detailed} and \textit{with examples}. As shown in Table  \ref{table:query_formulation_prompt}, \textit{simple} prompt refers to a prompt that only uses one sentence to briefly state the task for ChatGPT. This is seen as the most common usage for ChatGPT when users are not expert at constructing high-quality prompts. On the other hand, \textit{detailed} prompt means that a background story is included, which justifies clearly what is required for ChatGPT to successfully complete the task. The \textit{with example} prompt also includes an expected query formulation example, so that ChatGPT knows what is expected for it to generate a high-quality answer.

While the difference between the two \textit{detailed} prompts `q2' and `q3' is the way we describe the background story; for \textit{with example} prompts `q4' and `q5', `q5' also describes one Boolean query formulation strategy to identify PICO elements.  PICO refers to: Patient/ Problem, Intervention, Comparison and Outcome, and it is often used before systematic review search to clearly identify the research question and generate high-quality systematic review Boolean queries \cite{eriksen2018impact}. The intuition for including PICO element extraction in the prompt is that we want to help ChatGPT to better understand how the example query is formulated logically.

For each systematic review topic, we replace $"\{review\_title\}"$ with the title of the review. For prompts with an example, we further replace $"\{example\_review\_title\}"$ with the title of the example review topic and $"\{example\_review\_query\}"$ with the Boolean query used by that example review. For instance, for the example title \textit{"Thromboelastography (TEG) and rotational thromboelastometry (ROTEM) for trauma-induced coagulopathy in adult trauma patients with bleeding"}, we insert the Boolean query: 

{
\small
	(Thrombelastography[mesh:noexp] OR (thromboelasto*[All Fields] OR thrombelasto*[All Fields] OR ROTEM[All Fields] OR “tem international”[All Fields] OR (thromb*[All Fields] AND elastom*[All Fields]) OR (rotational[All Fields] AND thrombelast[All Fields])) OR (Thrombelastogra*[All Fields] OR Thromboelastogra*[All Fields] OR TEG[All Fields] OR haemoscope[All Fields] OR haemonetics[All Fields] OR (thromb*[All Fields] AND elastogra*[All Fields])))
}

\vspace{-10pt}
\subsection{Single Prompts for Query Refinement}

Single query refinement prompts refer to prompts that provide ChatGPT with a title of a systematic review and a corresponding Boolean query for that review, and instruct the model to produce a modification of that query that leads to higher search effectiveness. These prompts could be used by reviewers to improve a query they have manually devised, or could be integrated into a more complex pipeline of automation. 
We show our designed query refinement prompts in Table \ref{table:query_refinement_prompt}. 
 

\begin{table*}[t]
	\centering
	\small
	\begin{tabular}{p{3pt}cp{385pt}}
		\toprule
		&Prompt ID & Prompt \\
		\midrule
		\multirow{3}{*}{\rotatebox{90}{Simple}}
		&q6& For a systematic review seed Boolean query:  "\{initial\_query\}",  This query retrieves too many irrelevant documents and too few relevant documents about the information need: “\{review\_title\}”, Please correct this query so that it can retrieve fewer irrelevant documents and more relevant documents. \\
		 \midrule
		\multirow{3}{*}{\rotatebox{90}{With Example}}
		&q7& For a systematic review seed Boolean query:  “\{example\_review\_initial\_query\}"
		,This query retrieves too many irrelevant documents and too few relevant documents about the information need: “\{example\_review\_title\}”, therefore it should be corrected to: “\{example\_review\_refined\_query\}”. Now your task is to correct a systematic review Boolean query: "\{initial\_query\}" for information need “\{review\_title\}”, so it can retrieve fewer irrelevant documents and more relevant documents. \\ [5em]
		\bottomrule
	\end{tabular}
	
	\caption{ Prompts for single prompt query refinement}
	\vspace{-10pt}
	\label{table:query_refinement_prompt}
\end{table*}

We categorise prompts for query refinement into \textit{simple} and \textit{with example}. Similar to single prompt query formulation, \textit{simple} prompt `q6' indicates that the prompt is constructed only by briefly describing the task. For \textit{with example} prompt `q7', we include an example that tells ChatGPT what it means by successfully refining the query.
When asking ChatGPT to refine queries, we replace $"\{initial\_query\}"$ with the initial Boolean query we want ChatGPT to refine. For prompts that require an example, we further replace $\{example\_review\_title\}$ with the title of the example review topic; $"\{example\_review\_initial\_query\}"$ with the initial query of the example topic, and $"\{example\_review\_refined\_query\}"$ with the final query refined in the example topic. 


%
%

We investigate the effectiveness of two types of examples: (1) one high-quality systematic review example (HQE) and (2) an example that is similar or related to the querying topic (RE). To identify a related example, we use a monoBERT architecture to find the closest example from the test collection~\cite{nogueira2019multi}. Specifically, we concatenate the review title of the querying topic with the title of one potential example review and we pass it through PubMedBERT, a domain-specific BERT model pre-trained on PubMed abstracts \cite{pubmedbert}. We get a final classification score which refers to the relatedness of the potential example topic to the querying topic; we select the top-ranked example to include in the prompt during prompt creation.


\vspace{-10pt}
\subsection{Guided Prompts for Query Formulation}

We design a multi-step prompt that follows the same logic from one of the current state-of-the-art automated query formulation methods, namely the objective method~\cite{scells2020comparison}. We show our designed guided prompt with an example in Table \ref{table:designed_prompt}. Specifically, we follow a four step pipeline to generate the Boolean query: in the first step, Boolean query terms are identified using one seed study, then these terms are classified into four categories:  terms relating to health conditions (A), terms relating to a treatment (B), terms relating to types of study design (C), or others (N/A) in the second step. For the third step, terms in the same category are combined by `OR', and final queries are combined by `AND'. Finally, ChatGPT is asked to refine the query by adding more terms such as MeSH Terms. By doing a guided prompt, we can better control the terminology of Boolean query formulation from ChatGPT, also dividing the complex task into multiple, more manageable subtasks that are effective.

%
%

\begin{table*}
	\scriptsize
	\centering
	\begin{tabular}{lp{210pt}|p{195pt}}
		\toprule
		Step&Prompt & Example Answer\\
		\midrule
		Step1&Follow my instructions precisely to develop a highly effective Boolean query for a medical systematic review literature search. Do not explain or elaborate. Only respond with exactly what I request. First, Given the following statement and text from a relevant study, please identify 50 terms or phrases that are relevant. The terms you identify should be used to retrieve more relevant studies, so be careful that the terms you choose are not too broad. You are not allowed to have duplicates in your list. \newline statement: "Prevalence of Differentiated Thyroid Cancer in Autopsy Studies Over Six Decades: A Meta-Analysis" \newline Text:  Ten occult carcinomas of the thyroid gland were found in 274 unselected autopsies at the Pathology Service, Hospital José Joaquín Aguirre, between December 1980 and March 1983. This is the lowest incidence among the most recent published series. The smallest papillary microcarcinomas presented most frequently as nonencapsulated nonsclerosing tumors without inflammatory infiltrate or fibrosis, suggesting that they may represent the early stages of development. Our data show a relatively high and similar frequency of papillary microcarcinomas in surgical and autopsy series, but do not demonstrate risk factors for clinical evolution.&  1. Differentiated thyroid cancer \newline 2. Prevalence \newline 3. Autopsy studies \newline 4. Meta-analysis \newline 5. Occult carcinomas \newline 6. Thyroid gland \newline 7. Incidence \newline 8. Environmental factors \newline 9. Etiology \newline 10. Low risk \newline ...  \newline 50. Parameters. \\ \midrule

		Step 2&For each item in the list you created in step 1, classify it into as of three categories: terms relating to health conditions (A), terms relating to a treatment (B), terms relating to types of study design (C). When an item does not fit one of these categories, mark it as (N/A). Each item needs to be categorised into (A), (B), (C), or (N/A).  & 1. (A) Differentiated thyroid cancer \newline 2. (N/A) Prevalence \newline 3. (C) Autopsy studies \newline 4. (C) Meta-analysis \newline 5. (A) Occult carcinomas \newline 6. (A) Thyroid gland \newline 7. (N/A) Incidence \newline 8. (N/A) Environmental factors \newline 9. (N/A) Etiology \newline 10. (N/A) Low risk \newline ...  \newline 50. (N/A) Parameters.  \\ \midrule
		
		Step 3&Using the categorised list you created in step 2, create a Boolean query that can be submitted to PubMed which groups together items from each category. For example: ((itemA1[Title/Abstract] OR itemA2[Title/Abstract]  or itemA2[Title/Abstract]) AND (itemB1[Title/Abstract] OR itemB2[Title/Abstract] OR itemB3[Title/Abstract]) AND (itemC1[Title/Abstract] OR itemC2[Title/Abstract] OR itemC3[Title/Abstract])) & (((differentiated[Title/Abstract] OR thyroid[Title/Abstract] OR carcinoma[Title/Abstract] OR papillary[Title/Abstract] OR microcarcinoma[Title/Abstract]) AND (cancer[Title/Abstract] OR incidence[Title/Abstract] OR etiology[Title/Abstract] OR risk[Title/Abstract] OR gender[Title/Abstract] OR hormonal[Title/Abstract] OR nodular[Title/Abstract] OR goiter[Title/Abstract] OR Hashimoto's[Title/Abstract] OR malignancy[Title/Abstract] OR concomitant[Title/Abstract] OR tumor[Title/Abstract] OR infiltrate[Title/Abstract] OR fibrosis[Title/Abstract] OR development[Title/Abstract] OR frequency[Title/Abstract])) AND (autopsy[Title/Abstract] OR surgical[Title/Abstract] OR material[Title/Abstract] OR series[Title/Abstract] OR specimens[Title/Abstract] OR cases[Title/Abstract])). \\ \midrule
		
		Step 4&Use your expert knowledge to refine the query, making it retrieve as many relevant documents as possible while minimising the total number of documents retrieved. Also add relevant MeSH terms into the query where necessary, e.g., MeSHTerm[MeSH]. Retain the general structure of the query, however, with each main clause of the query corresponding to a PICO element. The final query still needs to be executable on PubMed, so it should be a valid query. & (((differentiated thyroid cancer[MeSH] OR "differentiated thyroid"[All Fields] OR "thyroid carcinoma"[All Fields] OR "papillary microcarcinoma"[All Fields]) AND (prevalence[All Fields] OR incidence[MeSH] OR "etiology of"[All Fields] OR "risk factors"[All Fields] OR gender[All Fields] OR hormonal[All Fields] OR "nodular goiter"[All Fields] OR "Hashimoto's thyroiditis"[MeSH] OR malignancy[MeSH] OR "concomitant lesion"[All Fields] OR tumor[All Fields] OR infiltrate[All Fields] OR fibrosis[All Fields] OR "early stages of development"[All Fields] OR frequency[All Fields])) AND (autopsy[MeSH] OR surgical[All Fields] OR material[All Fields] OR series[All Fields] OR specimens[All Fields] OR cases[All Fields]))  \\ 
		
		\bottomrule
		
	\end{tabular}
	\caption{Example designed guided prompt for query formulation.}
	\vspace{-20pt}
	\label{table:designed_prompt}
\end{table*}

\section{Experimental Settings}
Our experiments are conducted using two collections: CLEF technological assisted reviews (TAR) datasets \cite{kanoulas2017clef, kanoulas2018clef, kanoulas2019clef} and systematic review collection with seed studies (Seed Collection) \cite{wang2022little}. CLEF TAR dataset is published each year from 2017 to 2019 as a validation dataset for more effective systematic review Boolean query formulation and screening. We use CLEF TAR 2017 and 2018 as they both consist of only Diagnostic Test Accuracy systematic reviews. As CLEF TAR 2017 reviews is a subset of the CLEF TAR 2018 dataset, we obtained 72 review topics after removing duplicated reviews. Seed collection contains 40 systematic review topics with real seed studies that guide systematic review researchers during Boolean query formulation. Both collections contain systematic review titles, Boolean queries and relevant assessments.

To generate prompts, we replace fields in our designed prompts, shown in Table \ref{table:query_formulation_prompt}, \ref{table:query_refinement_prompt} and \ref{table:designed_prompt} with specific review topic information in the collection. For prompts with examples, we selected topic CD010438 from CLEF TAR collection as the HQE topic for generating queries for both collections, as the query obtained high effectiveness and has a simple structure for ChatGPT to follow.

We do not run query refinement for Seed collection as we could not get queries generated by objective and conceptual method reported in the original collection \cite{wang2022little}. 
The guided prompt method relies on the objective method, which requires at least one seed study to start with query formulation. Thus we did not run guided prompting on CLEF TAR as it does not contain seed studies. 


After obtaining generated Boolean queries from ChatGPT, any queries that produced incorrectly formatted Boolean queries were removed and generated again for the review topic.

To evaluate the generated Boolean queries, we executed them using PubMed's Entrez API to obtain retrieved PubMed IDs~\cite{canese2013pubmed}. We then used set-based measures such as precision, f-measure, and recall to assess the retrieved PubMed IDs using the abstract-level relevant assessment in the collection.

\section{Main Results}
\label{sec:results}

\subsection{Single Prompt Query Formulation}

\begin{table}
	\centering
	\small
	\begin{tabular}{p{2pt}p{30pt}c|cccc}
		\toprule
		
		&&Prompts&Precision&F1&F3&Recall\\ \midrule
		\multirow{10}{*}{\rotatebox{90}{CLEF}}&\multirow{3}{*}{Baselines}&Original&0.0207$^{*}$&0.0290$^{*}$&0.0481$^{*}$&0.8317$^{*}$\\ 
		&&Conceptual&0.0015$^{*}$&0.0027$^{*}$&0.0101$^{*}$&0.6997$^{*}$\\
		&&Objective&0.0002$^{*}$&0.0005$^{*}$&0.0023$^{*}$&\textbf{0.9128}$^{*}$\\

		\cmidrule{2-7}
		&\multirow{1}{*}{Simple}&q1&0.0543&0.0500&0.0590&0.1293$^{*}$\\\cmidrule{2-7}
		&\multirow{2}{*}{Detailed}&q2&\textbf{0.1166}&0.0654&0.0696&0.1310$^{*}$ \\ 
		&&q3&0.0844&0.0443&0.0497$^{*}$&0.1175$^{*}$\\\cmidrule{2-7}
		&\multirow{2}{*}{With Example}&q4&0.0752&0.0642&\textbf{0.0847}&0.5035 \\
		&&q5&0.0958&\textbf{0.0717}&0.0844&0.3335$^{*}$ \\ \midrule
		\midrule
		
		\multirow{10}{*}{\rotatebox{90}{SC}}&\multirow{3}{*}{Baselines}&Original&0.0367&\textbf{0.0651}$^{*}$&\textbf{0.1099}$^{*}$&\textbf{0.7366}$^{*}$ \\
		&&Conceptual&0.0018&&&0.4138\\
		&&Objective&0.0057&&&0.5192 \\
		
		\cmidrule{2-7}
		&\multirow{1}{*}{Simple}&q1&0.0501&0.0274&0.0298&0.0528 \\\cmidrule{2-7}
		
		&\multirow{2}{*}{Detailed}&q2&\textbf{0.0983}&0.0310&0.0278&0.0394$^{*}$\\ 
		&&q3&0.0730&0.0329&0.0329&0.0519 \\ \cmidrule{2-7}

		&\multirow{2}{*}{With Example}&q4&0.0283&0.0274&0.0374&0.1290 \\
		&&q5& 0.0188&0.0193&0.0271&0.0785\\ 
		
		\bottomrule
	\end{tabular}
	\caption{Single Prompt query formulation results. $CLEF$ indicates CLEF TAR collection and $SC$ indicates seed collection. Statistical significant differences (Student's two-tailed, paired t-test with Bonferonni correction, p < 0.05) between q4 and all other methods are indicated by $^{*}$.}
	\vspace{-10pt}
	\label{table:query_formulation_result}
\end{table}
	
We show results of single prompt query formulation in Table \ref{table:query_formulation_result}.
The results indicate that queries generated from ChatGPT generally obtain a higher precision compared to the current-state-of-the-art automatic query formulation methods, with a trade off of lower recall. 
For F-measure (which captures both precision and recall), ChatGPT generated queries are more effective than both state-of-the-art and original authored queries on the CLEF collection. However, they are less effective on the Seed Collection.
Search in systematic reviews generally requires high-recall to ensure that all evidence can be found relating to the research topic. All the ChatGPT generated queries obtain a lower recall than the baseline methods, suggesting that ChatGPT generated queries may be not suitable for high-recall retrieval, but rather best suited when time is limited; e.g., for rapid reviews \cite{marshall2019rapid}.

Using different Simple and Detailed prompts (q1--3) only had a minor impact on effectiveness. For CLEF, q2 statistically significantly better with respect to precision; otherwise the prompt type did not have a strong effect.
However, we found that prompts that include a high quality systematic review topic as example are able to significantly outperform those  without, shown as a consistently higher F\_1, F\_3 and recall. When comparing the effectiveness of two prompts with examples, we found that asking ChatGPT to generate PICO elements before generating its final Boolean query resulted in Boolean queries with a lower recall but higher precision. Overall, our findings indicate that including a high quality systematic review query example in the prompt is crucial, while the level of detail in the task description may not have a significant impact.

\begin{table}
	\centering
	\small
	\begin{tabular}{p{1.5pt}p{40pt}|p{60pt}p{60pt}p{60pt}p{60pt}}
		\toprule
		&&Precision&F1&F3&Recall\\ \midrule
		\multirow{3}{*}{\rotatebox{90}{CLEF}}&q4-HQE&0.0751&0.0642&0.0872&0.5035\\ \cmidrule{2-6}
		&q4-RE& 0.1105($+47.1\%$)&0.0909($+41.6\%$)&0.1144($+31.2\%$)&0.4183($-38.1\%$)\\ \midrule
		
		\multirow{3}{*}{\rotatebox{90}{SC}}&q4-HQE&0.0283&0.0274&0.0374&0.129\\ \cmidrule{2-6}
		&q4-RE& 0.0351($+24.0\%$)&0.0140($-48.9\%$)&0.0139($-62.8\%$)&0.0161$^{*}$($-87.6\%$)\\ 
		
		\bottomrule
	\end{tabular}
	\caption{Comparison of result for single prompt query generation prompt `q4' when using different type of examples.  $CLEF$ indicates CLEF TAR collection and $SC$ indicates seed collection; For each collection, two type of example is used, $q4-HQE$ refers to using one high quality example, while $q4-RE$  refers to using related query as example. Statistical significant differences (p < 0.05) between the two types of examples are indicated by $^{*}$.}
	\label{table:example_comparison_query_formulation_result}
		\vspace{-30pt}
\end{table}

Next, to assess the impact of query construction example to the effectiveness of generated Boolean queries using ChatGPT, we further test the effectiveness when different types of examples is used, as described in section \ref{single_p_query_generation}.

 Table \ref{table:example_comparison_query_formulation_result} compares the effectiveness of queries generated using the most relevant topic in the prompt to queries generated using a high-quality example. Using a relevant topic as example can result in queries with higher precision, but lower recall.





\begin{figure*}[t!]
	\begin{subfigure}[t!]{.33\textwidth}
		\includegraphics[width=\textwidth]{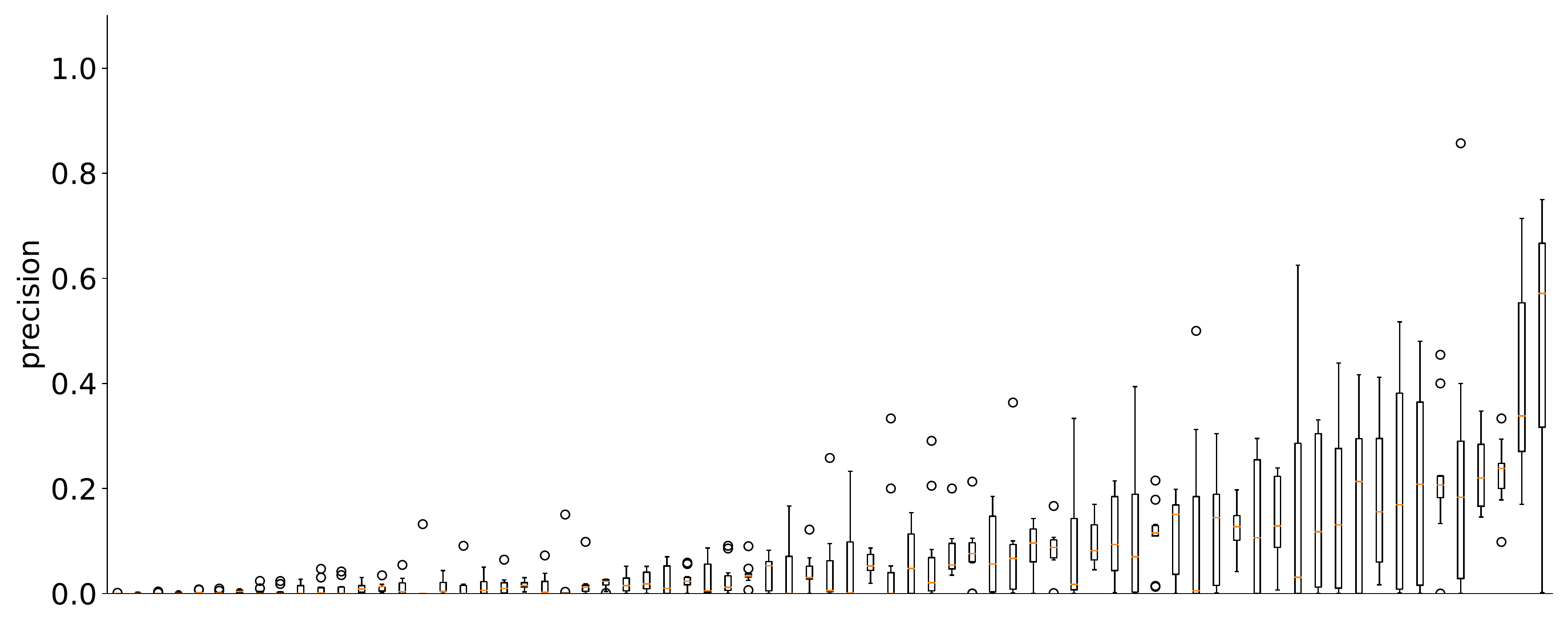}
		\caption{CLEF Precision}
		\label{fig:results.2017var-single}
	\end{subfigure}
	\begin{subfigure}[t!]{.33\textwidth}
		\includegraphics[width=\textwidth]{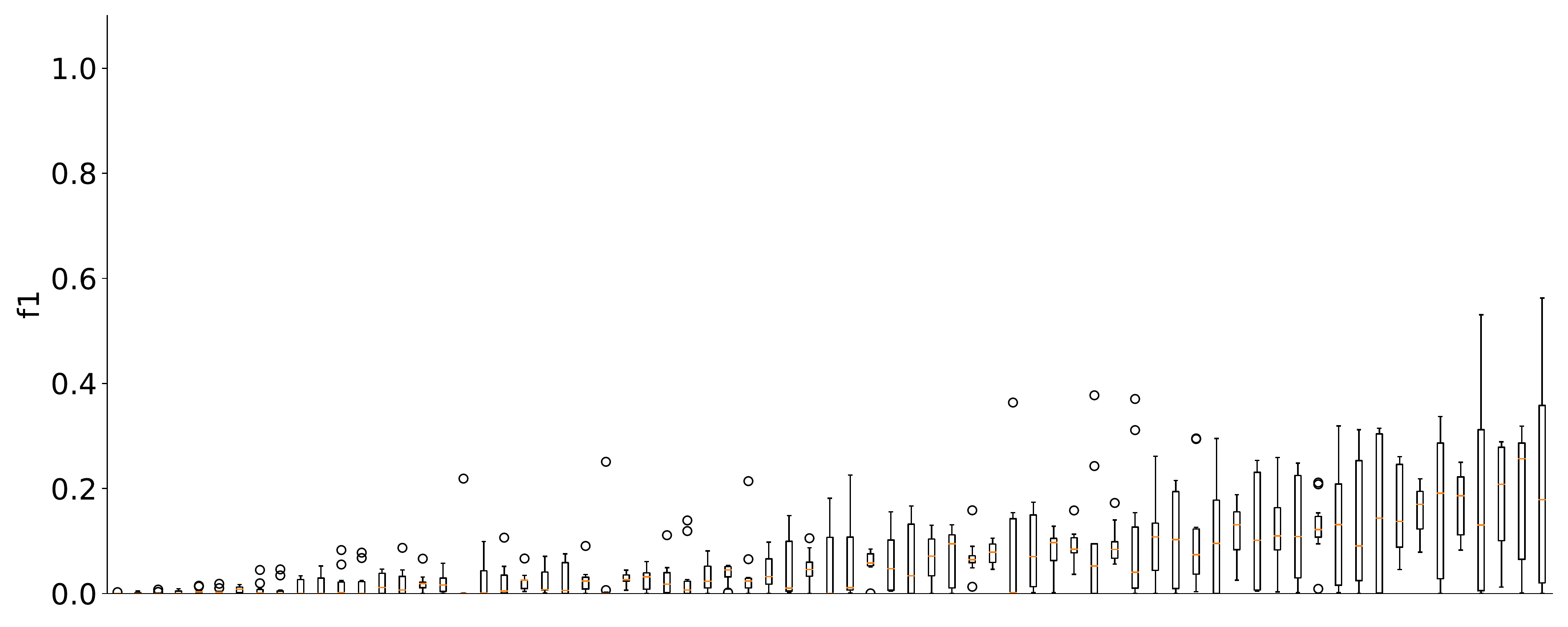}
		\caption{CLEF F1}
		\label{fig:results.2018var-single}
	\end{subfigure}
	\begin{subfigure}[t!]{.33\textwidth}
		\includegraphics[width=\textwidth]{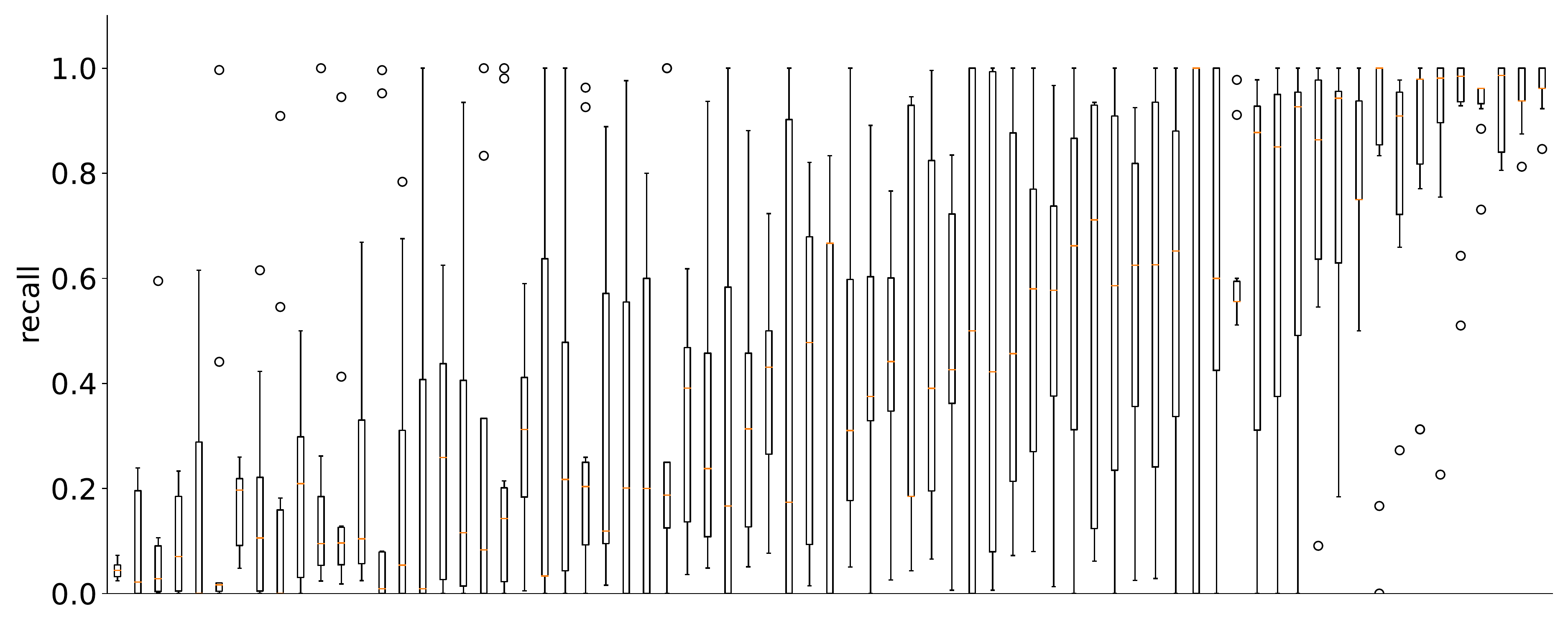}
		\caption{CLEF Recall}
		\label{fig:results.2018var-multi}
	\end{subfigure}	
	
		\begin{subfigure}[t!]{.33\textwidth}
		\includegraphics[width=\textwidth]{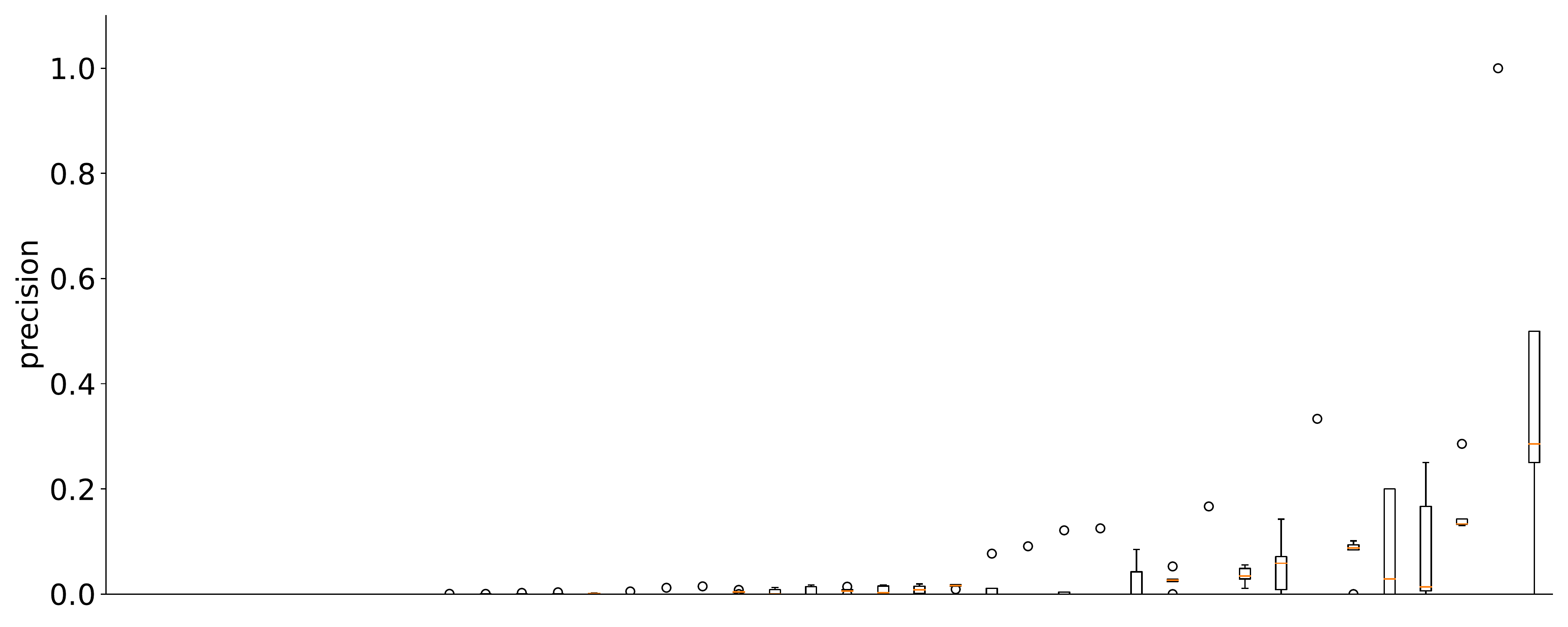}
		\caption{SC Precision}

	\end{subfigure}
	\begin{subfigure}[t!]{.33\textwidth}
		\includegraphics[width=\textwidth]{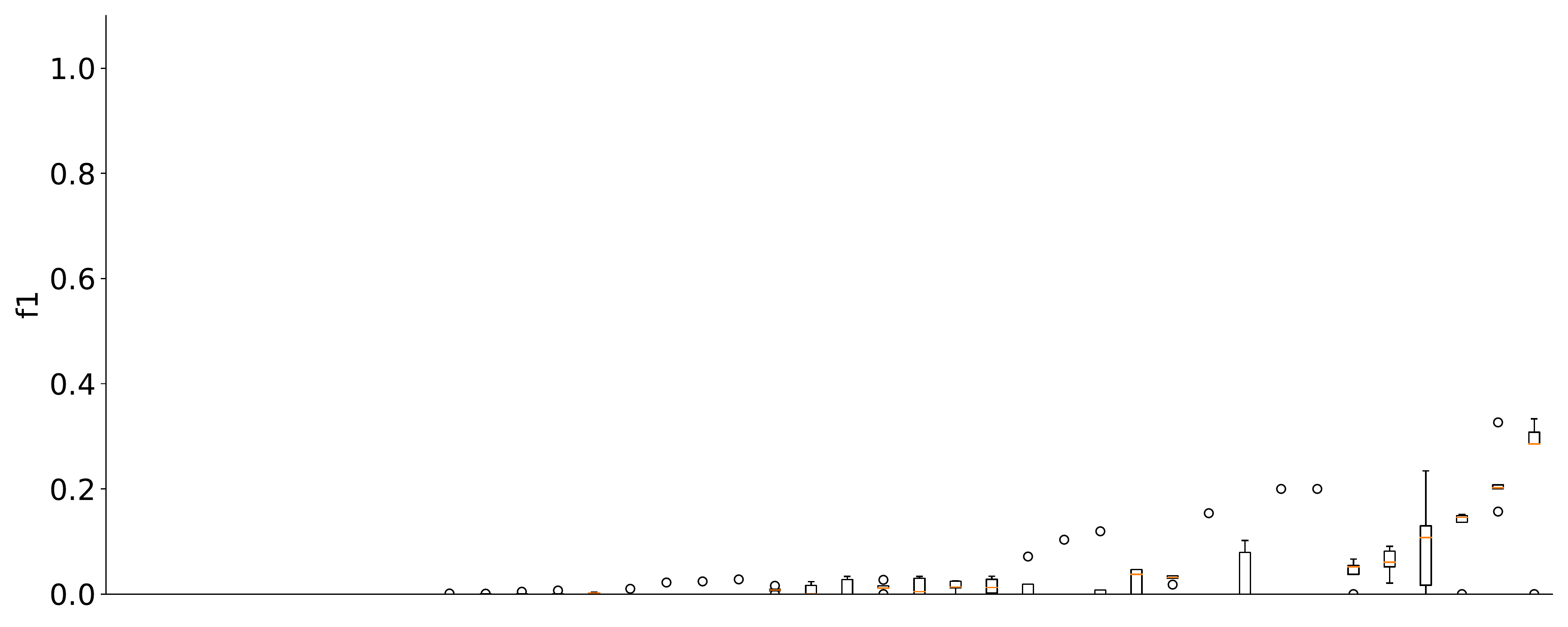}
		\caption{SC F1}

	\end{subfigure}
	\begin{subfigure}[t!]{.33\textwidth}
		\includegraphics[width=\textwidth]{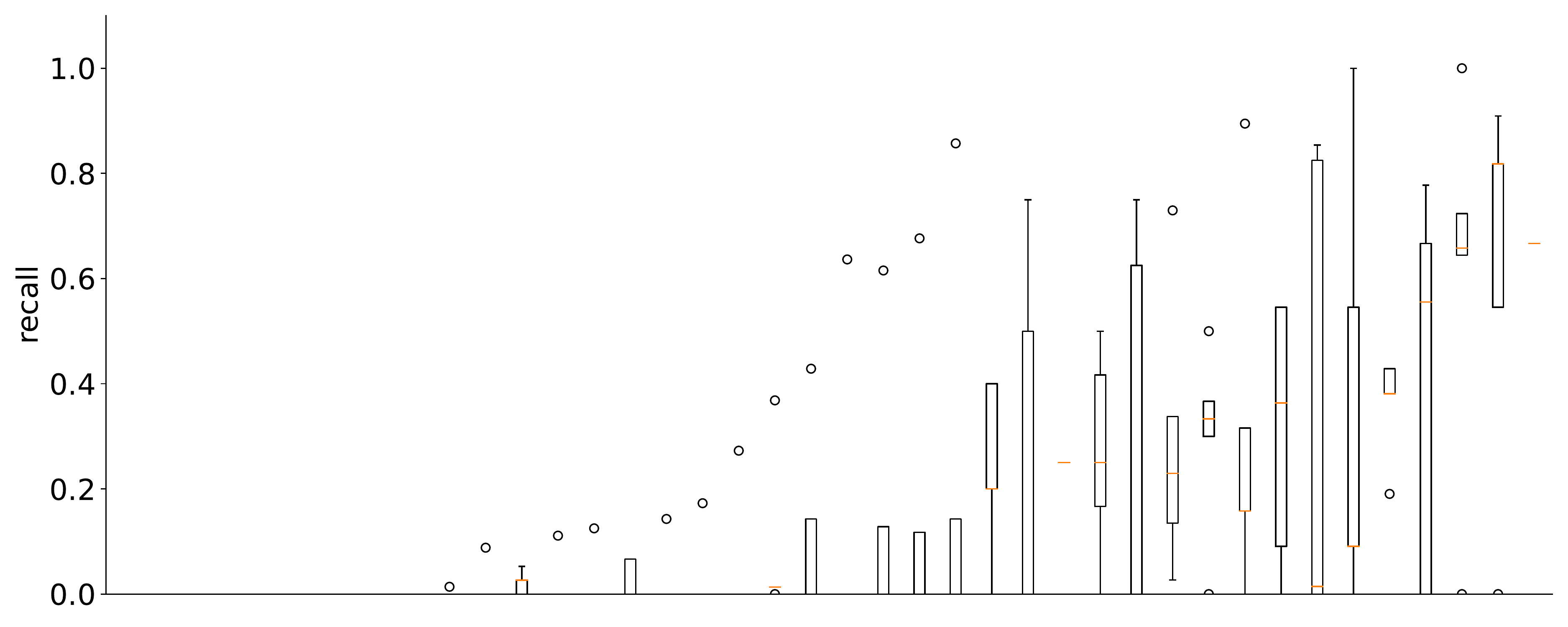}
		\caption{SC Recall}

	\end{subfigure}	

\vspace{-10pt}
	
	\caption{Topic-by-topic variability boxplot for effectiveness of 10 iterative runs in single prompt query formulation. $CLEF$ indicates CLEF TAR collection and $SC$ indicates seed collection. }
	
	\label{fig:variability_formulation}
		\vspace{-10pt}
\end{figure*}

\cgpt will generate different responses for the same prompt. To study variability of effectiveness, we select the more effective prompt, q4, and we re-run the prompt 10 times.
Variability is shown in Figure \ref{fig:variability_formulation}. Recall varied more than precision and F-measures: variance of recall is 12\% of its mean value; precision was 7.1\%, F-1 6.6\% and F-3 7.2\%. Our result indicates that the generated queries from the same prompt would mainly differ in the ability to obtain more relevant documents (recall) from the included studies.

\vspace{-10pt}
\subsection{Single Prompt Query Refinement}

The results of single prompt query refinement, as shown in Table \ref{table:query_refinement_result}, indicate that ChatGPT is capable of improving the effectiveness of systematic review Boolean queries. Specifically, the use of ChatGPT for query refinement leads to an increase in precision and F-measure, while obtaining a lower recall. Therefore, it is crucial to first create a seed query with a high recall, and then use ChatGPT to refine the query in order to achieve highly effective Boolean queries. 

Furthermore, our results indicate that the greatest improvement in effectiveness is achieved when using seed queries that are automatically formulated by the objective method. This will also result in query with the highest recall among all methods. After refinement using ChatGPT, there is only a 11\% drop in recall, with considerable gains in precision, F-1, and F-3.

\begin{figure*}[t!]
	\begin{subfigure}[t!]{.33\textwidth}
		\includegraphics[width=\textwidth]{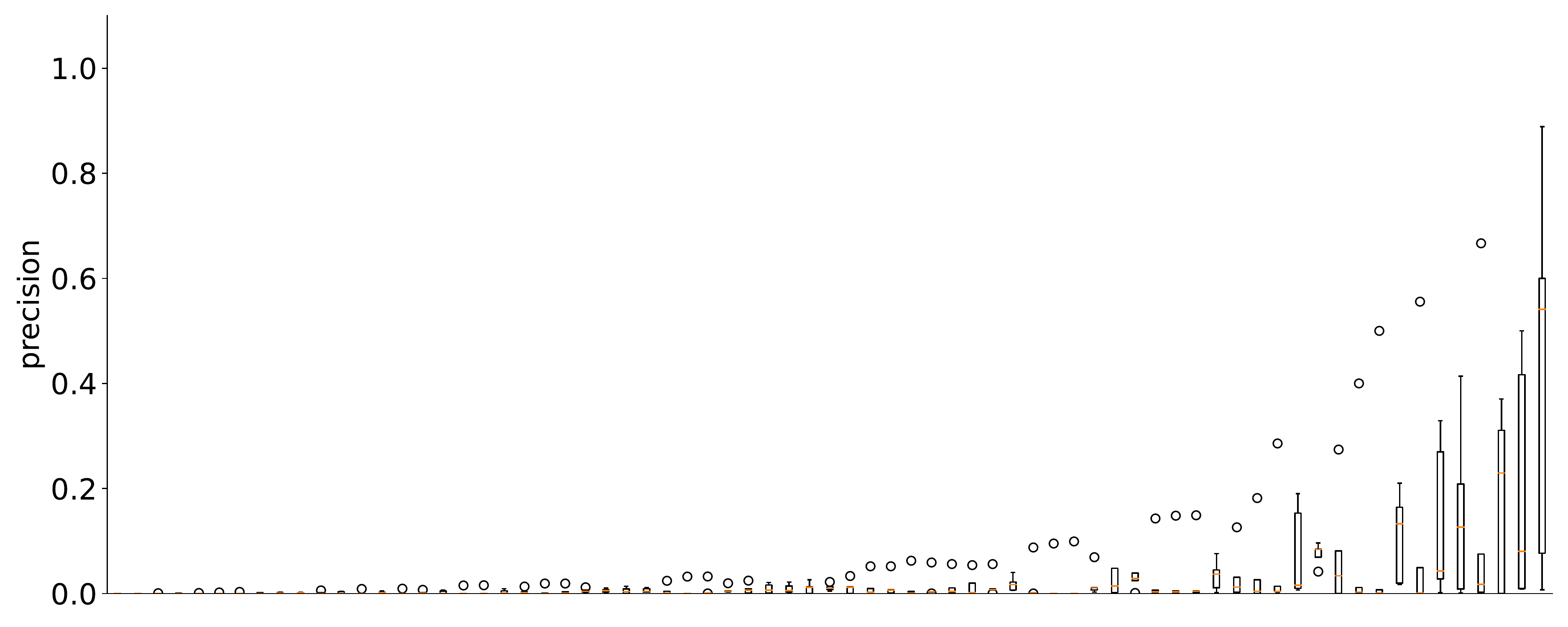}
		\caption{Precision}
		\label{fig:precision-refinement}
	\end{subfigure}
	\begin{subfigure}[t!]{.33\textwidth}
		\includegraphics[width=\textwidth]{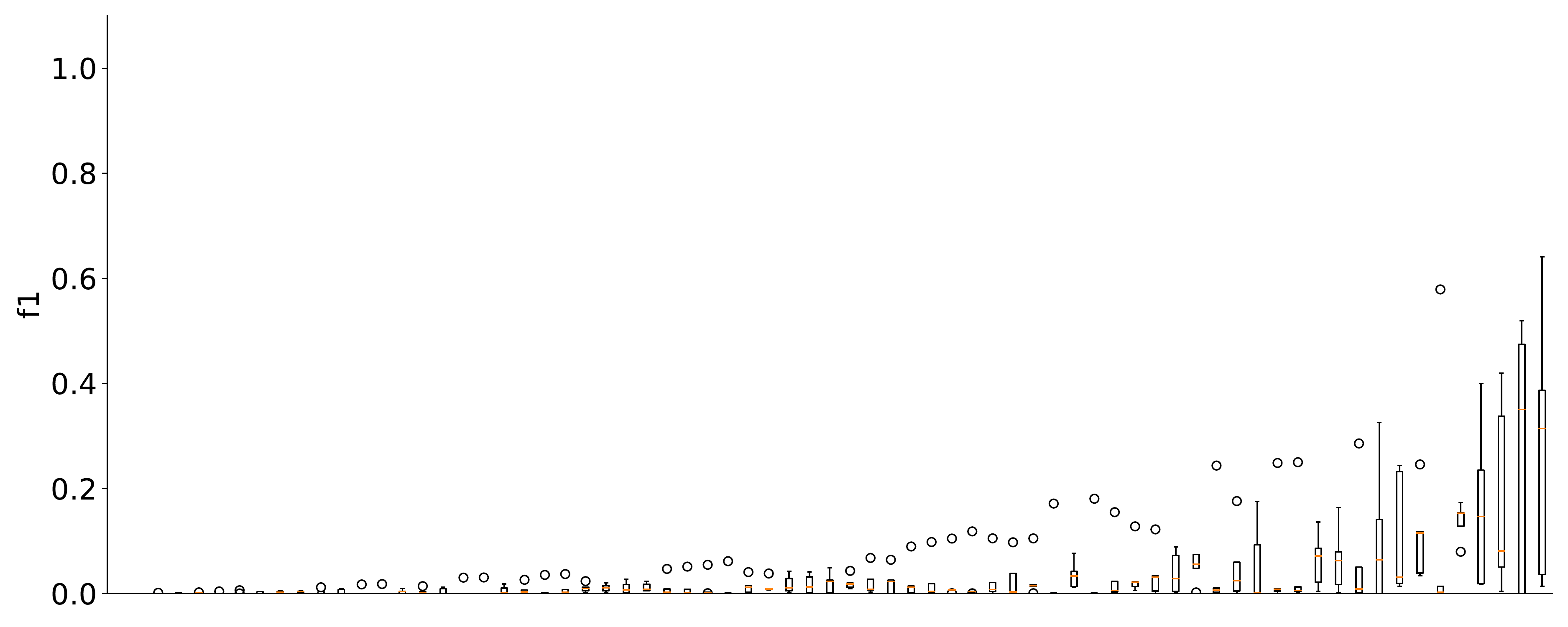}
		\caption{F1}
		\label{fig:f1_refinement}
	\end{subfigure}
	\begin{subfigure}[t!]{.33\textwidth}
		\includegraphics[width=\textwidth]{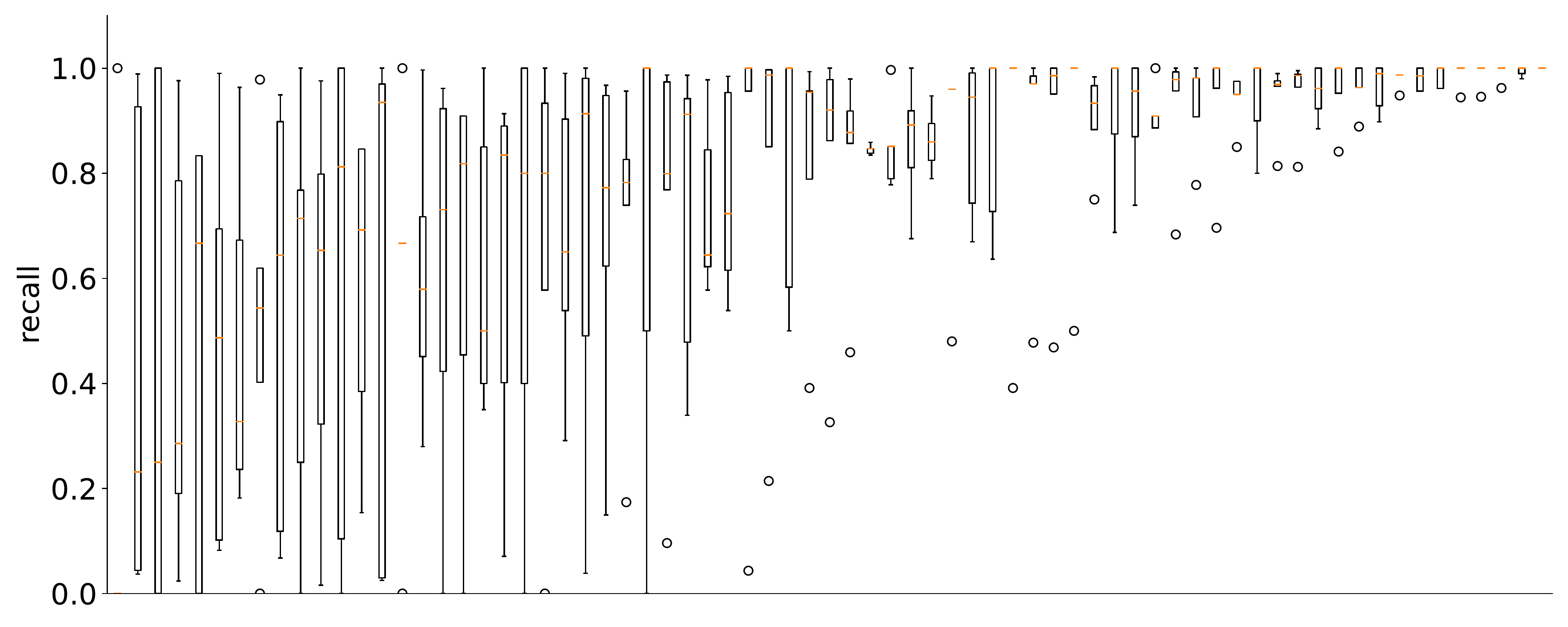}
		\caption{Recall}
		\label{fig:recall-refinement}
	\end{subfigure}	
	
	\caption{Topic-by-topic variability boxplot for effectiveness of 10 iterative runs in single prompt query refinement.}
	\label{fig:variability_refinement}
\end{figure*}

Variability is once again studies by running the best method, refined-objective run, 10 times. We show the variability of the query refinement in Figure \ref{fig:variability_refinement}. 
There is less variance in query refinement than query formation (Figure~\ref{fig:variability_formulation}). This is understandable given the query structure is already provided by the seed query, whereas query formulation must be done from scratch from the title of the review.

\begin{table}
	\centering
	\small
	\begin{tabular}{l|p{32pt}p{32pt}p{32pt}p{32pt}}
		\toprule
		
		Prompts&Precision&F1&F3&Recall\\ \midrule

		Original&0.0207&0.0290&0.0481&0.8317 \\
		q6-Original&0.0795$^{*}$&0.0597$^{*}$&0.0802$^{*}$&0.5060$^{*}$\\ \midrule
		
		Conceptual&0.0014&0.0027&0.0100&0.6996\\
		q7-conceptual&0.0022&0.0039&0.0069&0.2699$^{*}$\\ \midrule

		Objective&0.0002&0.0005&0.0023&\textbf{0.9128} \\
		q7-Objective&0.0460$^{*}$&0.0471$^{*}$&0.0652$^{*}$&0.8115$^{*}$\\ \midrule
		
		q4&0.0751&0.0642&0.0872&0.5035 \\
		q7-q4&\textbf{0.1162}&\textbf{0.0772}&\textbf{0.0921}&0.3179$^{*}$\\
		
		\bottomrule
	\end{tabular}
	\caption{Result table for Single Prompt query refinement on CLEF TAR collection. For a refinement method `q6-Original', `q6' indicates the prompt used to generate the refined query; `Original' indicate the seed queries used for ChatGPT to refine. For each query refinement method, statistical significant differences (p < 0.05) between refined prompt and seed queries is indicated by $^{*}$.}
	\label{table:query_refinement_result}
		\vspace{-10pt}
\end{table}

\subsection{Guided Prompt Query Formulation}

The result of guided prompt query formulation from Table \ref{table:designed_prompt_result} shows that if using well-chosen seed study, queries generated from guided prompt are more effective than single prompt queries. 

\begin{figure*}[t!]
	\begin{subfigure}[t!]{.33\textwidth}
		\includegraphics[width=\textwidth]{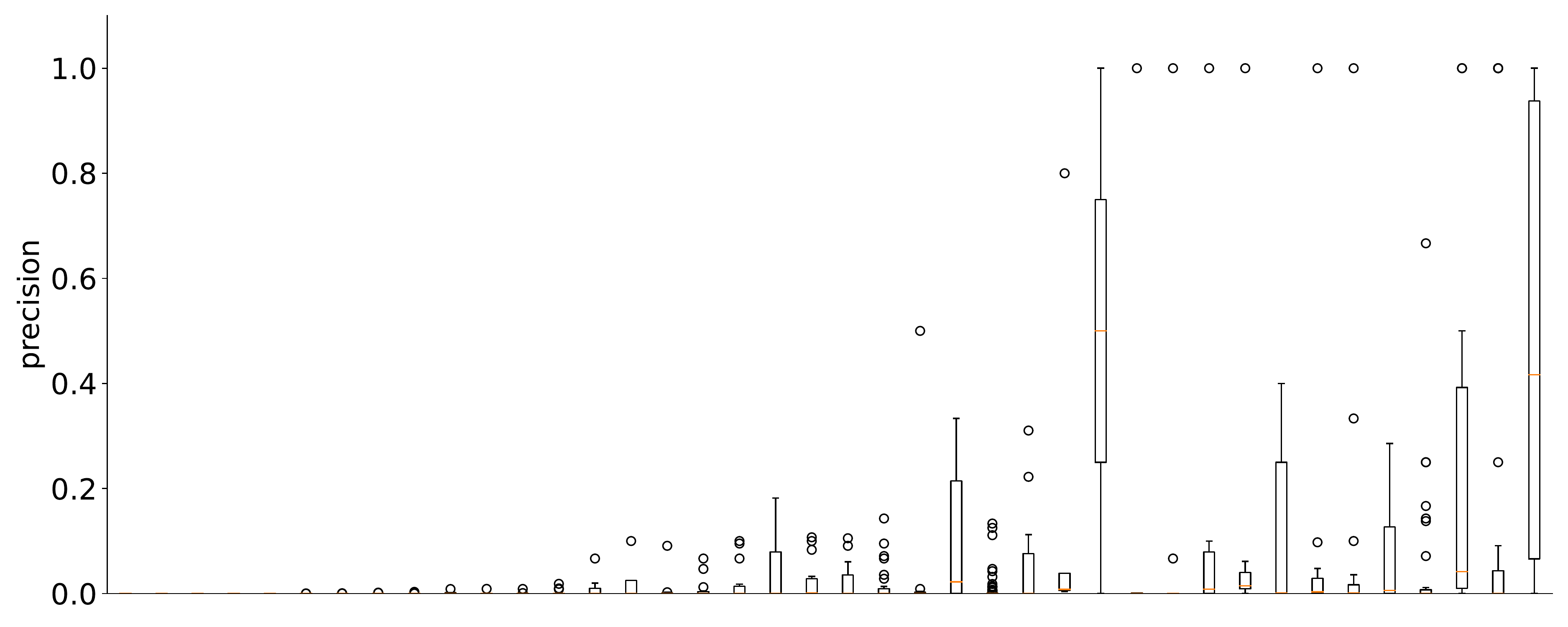}
		\caption{Precision}
		\label{fig:precision-guided}
	\end{subfigure}
	\begin{subfigure}[t!]{.33\textwidth}
		\includegraphics[width=\textwidth]{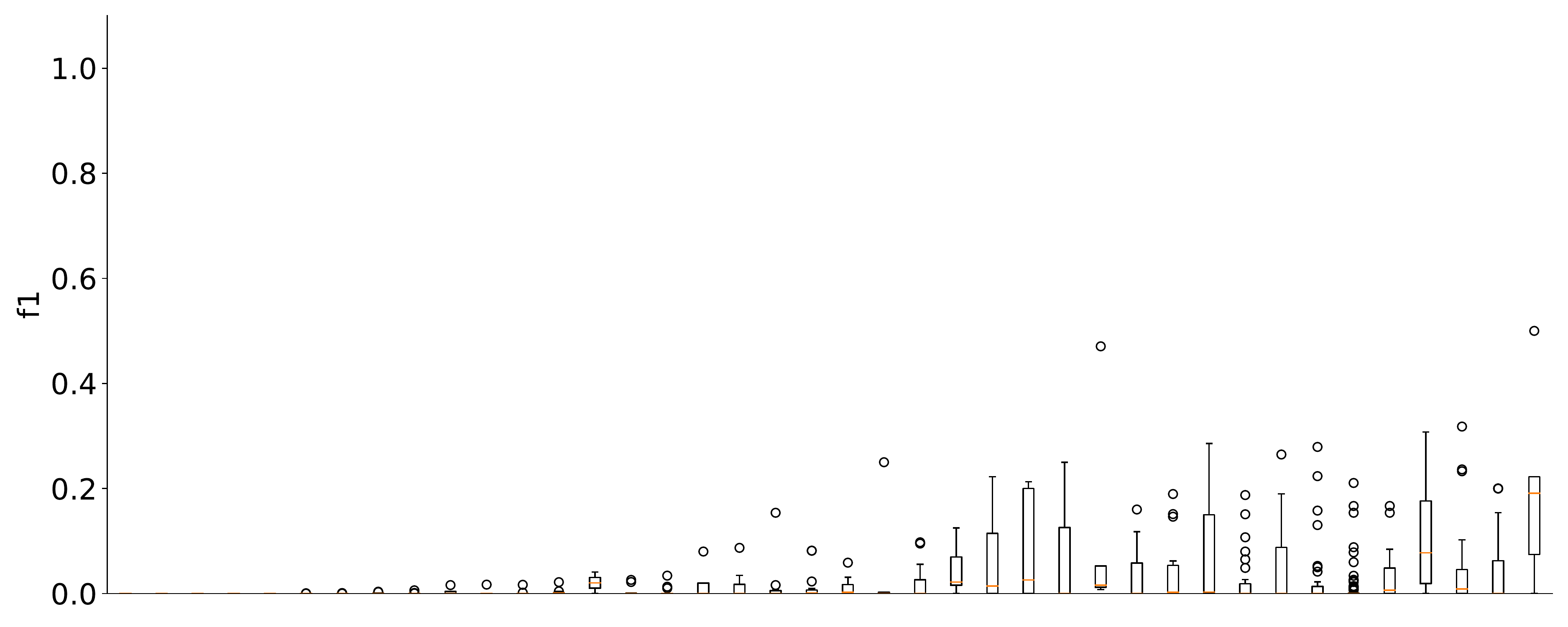}
		\caption{F1}
		\label{fig:f1_guided}
	\end{subfigure}
	\begin{subfigure}[t!]{.33\textwidth}
		\includegraphics[width=\textwidth]{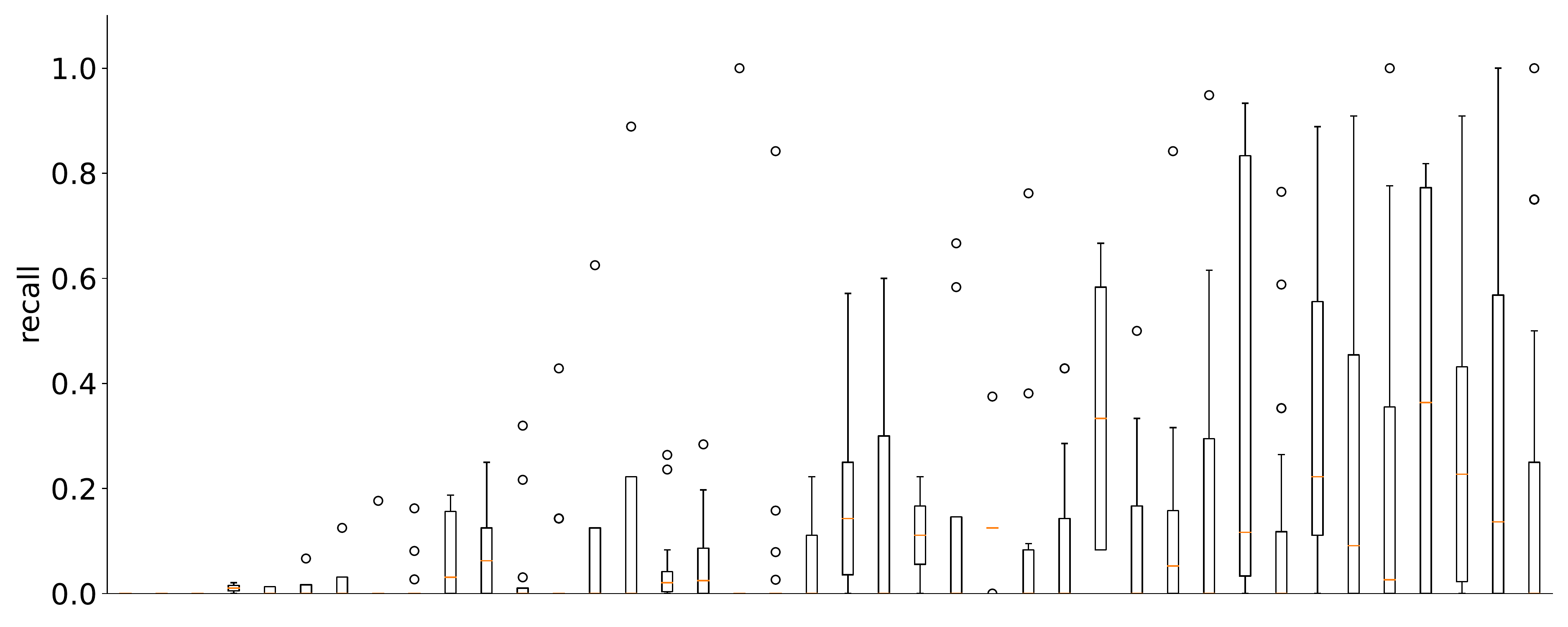}
		\caption{Recall}
		\label{fig:recall-guided}
	\end{subfigure}	
	
	\caption{Topic-by-topic variability boxplot for effectiveness of using different seed studies for guided prompt query formulation.}
	\label{fig:variability_guided}
\end{figure*}

\begin{figure*}[t!]
	\begin{subfigure}[t!]{.33\textwidth}
		\includegraphics[width=\textwidth]{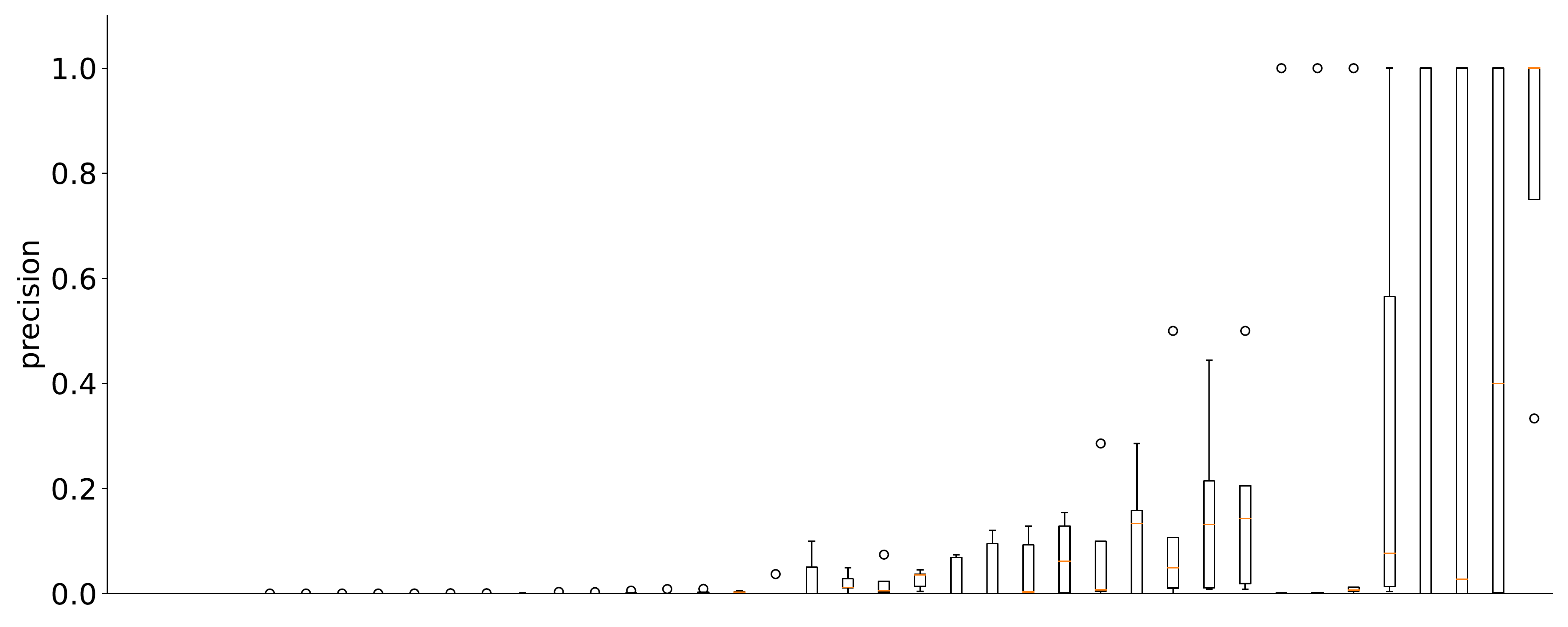}
		\caption{Precision}
		\label{fig:precision-guided-s}
	\end{subfigure}
	\begin{subfigure}[t!]{.33\textwidth}
		\includegraphics[width=\textwidth]{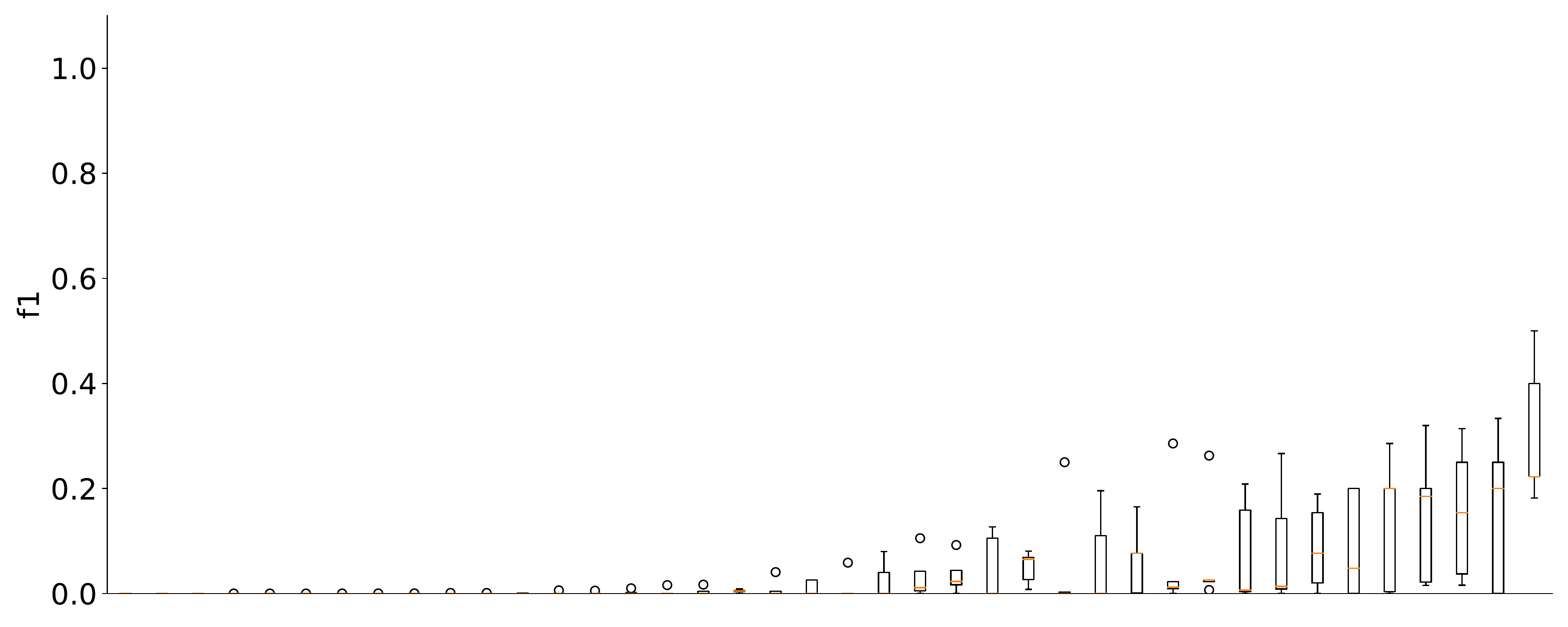}
		\caption{F1}
		\label{fig:f1_guided-s}
	\end{subfigure}
	\begin{subfigure}[t!]{.33\textwidth}
		\includegraphics[width=\textwidth]{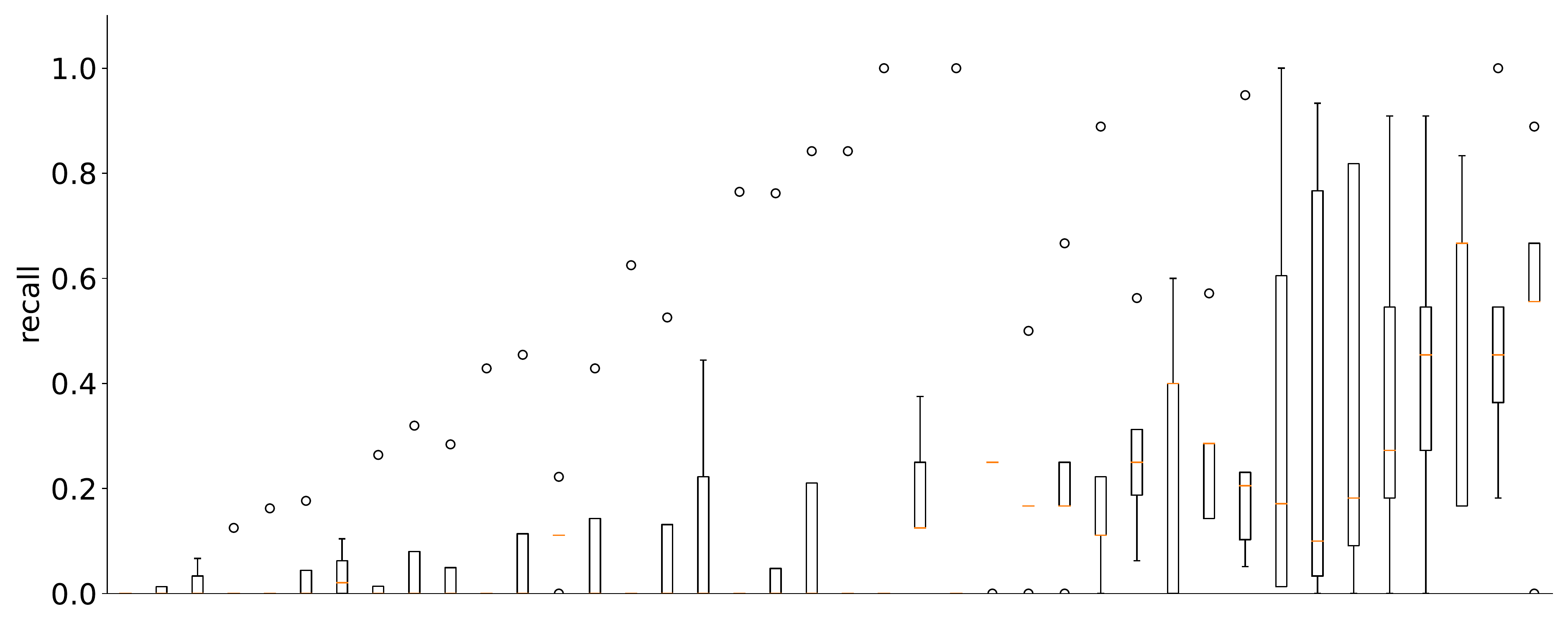}
		\caption{Recall}
		\label{fig:recall-guided-s}
	\end{subfigure}	
	
	\caption{Topic-by-topic variability boxplot for effectiveness of 10 iterative runs using the same seed study in guided prompt query formulation.}
	\label{fig:variability_guided-s}
\end{figure*}

However, like previous experiments in single prompt query generation and refinement, the effectiveness varies considerably across runs; and the effectiveness also depends on the seed study being used. We show the variability of query effectiveness in Figure \ref{fig:variability_guided} when different seed studies are used to generate queries, and in Figure \ref{fig:variability_guided-s}, we show the variability of effectiveness when the same seed study is used, picked from the best seed study effectiveness from the first run.

From the variability graph, we see that query generation using guided prompt is not stable across different seed studies. Furthermore, even when same seed study is used to generate multiple runs, there is a high degree of variability. The range of precision and recall for some topics can span from 0 to 1, especially when the average effectiveness is high.

\begin{table}
	\centering
	\small
	\begin{tabular}{l|p{60pt}p{60pt}p{60pt}p{60pt}}
		\toprule
		
		Prompts&Precision&F1&F3&Recall\\\midrule
		
		q4&0.0284&0.0274&0.0374&0.1290 \\ \midrule
		
		Guided&0.0993(+249.6\%)&0.0492(+79.6\%)&0.0565(+51.1\%)&0.5171(+301.6\%)$^{*}$ \\	
		\bottomrule
	\end{tabular}
	\caption{Result table for Guided Prompt query formulation on Seed Collection, compared with single prompt query generation `q4'; Statistical significant differences (p < 0.05) between guided prompt and single prompt is indicated by $^{*}$. }
	\label{table:designed_prompt_result}
\end{table}

\section{Query Failure Analysis}
\label{sec:failings}

We attempt to uncover the characteristics of queries with poor effectiveness, with a guise of identifying ways to improve query generation. To do this, we follow the steps below to select a set of successful and failing queries:
\begin{enumerate}[leftmargin=15pt,noitemsep,topsep=0pt]
	
	\item Select the best method for query formulation (q4) and query refinement (q7-objective).
	
	
	\item From the 10 iterative runs from each task, we extract an `oracle' result, using recall as an indicator of effectiveness. (Tie broken using precision.)


	\item Next we compare the oracle effectiveness with that of the original query. Successful topics are defined as those where the oracle has higher precision and recall than the original query; failing topics are those where the oracle has lower precision and recall than the original query. The intuition of selecting successful topics and failing topics above is based on the fact that higher recall often means more documents are retrieved from the query, and results in lower precision.
	
\end{enumerate}
Only the CLEF dataset was used because the Seed Collection contains topics that are not necessarily systematic reviews; e.g., scoping review, rapid review, etc.




Using this method above, we identify seven successful topics\footnote{Topics CD008759, CD009323, CD007427, CD010023, CD010213, CD011515, CD008686} and six failed topics \footnote{Topics CD010705, CD009591, CD010680, CD010502, CD010296 and CD011126.} for query formulation. For query refinement, we identify three succesfull topics \footnote{Topics CD010023, CD010213, CD007427}, and eleven topics for query refinement \footnote{Topics CD010657, CD012165, CD009135, CD011926, CD008643, CD011134, CD012083, CD008782, CD009579, CD008691, CD012010.}.


By comparing the failed queries with other queries, we summarize three key findings:

\begin{enumerate}[leftmargin=10pt,noitemsep,topsep=0pt]
	\item Poorly performing ChatGPT generated queries tended to retrieve a large number of results: for comparison of successful queries and failing queries for query formulation, the median ratio for successful queries is  0.30, while for failing queries, this ratio is 10.28. Same for query refinement, the median ratio for or successful queries and failing queries are 1.39 and 2.74, respectively. This means that failing queries may be categorised by having a large number of OR clauses; thus it may be possible to detect such queries and introduce mitigation strategies. 

	\item We also notice that some of the ChatGPT queries contain many incorrect MeSH Terms. For the best performing prompt for query formulation, on average three MeSH Terms were generated, and 55\% of the MeSH Terms generated were not in the MeSH vocabulary. On the other hand, for query refinement prompt, only 1.5 MeSH Terms were generated on average per query, with 66\% of the MeSH Terms not in the MeSH vocabulary. The is no strong correlation between the ratio of incorrectly generated MeSH Terms and query effectiveness. However, we also acknowledge that this does not mean certain incorrect MeSH Terms do not have a major negative impact on effectiveness. Further experiment is needed (e.g., MeSH Term correction \cite{wang2021mesh, wang2022automated}) to check how these MeSH Terms impact query effectiveness and how would effectiveness changes if MeSH Terns are corrected.
	
	\item Relevance judgements were done using the original query so may be biased towards this. In all the failing queries, the number of unjudged document in the retrieved documents set accounts for more than 94.8\% of the retrieved documents, much higher than that from the original queries 63.7\%; On the other hand, the portion of unjudged documents from generated and original queries for other topics was similar, at 56.2\% and 55.4\%, respectively. It may be that the failing queries simply retrieved many more unjudged documents because they had poor performance. But there is also a possibility that they actually managed to find clusters of documents relevant to topics but never assessed by reviewers. While this is conjecture, it does spark some ideas for solutions to improve such queries. If there are, in fact, clusters of relevant documents amongst a large number of non-relevant ones, then automated classification techniques may help to uncover these~\cite{marshall2018machine,timsina2016advanced,kim2011automatic,stansfield2013clustering,adeva2014automatic}. In particular, active learning~\cite{cohen2006reducing,miwa2014reducing} may be used to selectively assess documents from different clusters to hone in on relevant documents.
	
\end{enumerate}


\section{Summary of Findings}
In light of the empirical results reported in Section~\ref{sec:results}, we provide a summary of the key findings of our study in terms of the research questions we set to investigate.

\par \textbf{RQ1:} \textit{How does ChatGPT compare with current state-of-the-art methods for formulating and refining systematic review Boolean queries?}

In terms of automatic query formulation, our results indicate that the use of ChatGPT compares favourably with the current state-of-the-art automated Boolean query generation methods in terms of precision, at the expenses of a lower recall.

In terms of automatic query refinement, our results indicate that ChatGPT is effective at refining existing Boolean queries that have been generated by a previous step of automated Boolean query formulation. 
In particular, the ChatGPT-based refinement of queries generated using the objective method for Boolean query formulation leads to the most considerable improvements in effectiveness.

In both tasks, improvements are observed in terms of increased precision, i.e. a reduction of irrelevant documents being retrieved by the query. These improvements come at the expense of losses in recall (though for the best query refinement settings these losses are only marginal). High recall is often a key requirement for traditional systematic reviews and thus ChatGPT may not be well suited in these settings. However, we note that typically, in manually created Boolean queries, MeSH terms are used to improve recall, but the queries obtained through ChatGPT contain a large number of incorrect MeSH terms (see Section~\ref{sec:failings}). The addition of a post-processing step that resolves incorrect MeSH terms produced by ChatGPT may alleviate the losses in recall we witnessed in the results. For this, it may be possible to adapt existing methods for MeSH term suggestion~\cite{wang2021mesh,wang2022automated}, which also could be used to further refine the queries produced by ChatGPT by adding more effective MeSH terms the model may have failed to identify. We further note that within the systematic review process, the step of snowballing is designed to further increase recall. Snowballing refers to the recursive analysis of references cited in retrieved documents~\cite{greenhalgh2005effectiveness} -- in all effects, this process adds to the set to be reviewed studies not retrieved by the Boolean query. Thus, it may be that losses in recall observed when using ChatGPT to create queries may be recovered through snowballing on the identified relevant studies.


\par \textbf{RQ2:} \textit{To what extent do the prompts used to generate systematic review Boolean queries impact the effectiveness of the Boolean queries produced by ChatGPT?}

Our results suggest that the type of prompt used has considerable effects on the effectiveness of the queries produced by ChatGPT. We further review the impact of guided prompts in the next research question, and here we only focus on single prompts. 

In this case, we observe that integrating example Boolean queries into the prompts tends to provide better queries than simple prompts (Table~\ref{table:query_formulation_result}). This is particularly the case for the CLEF datasets -- while the same does not hold on the seed collection, we note this collection is limited in the number of topics available, so results may be less generalisible. Notably, improvements obtained when including query examples are large for recall (this is also the case on the seed collection), thus partially mitigating the low recall observed throughout the queries produced with ChatGPT. We also observe that the nature of the example query influences results. Specifically, example queries that are semantically close to the systematic review topic for which the new query is being created lead to higher precision. On the other hand, using high quality but potentially unrelated queries in the prompts, instead, leads to higher recall. 

We also have observed that the inclusion in the prompt of instructions to generate PICO elements does lead to considerable losses in recall, and provides improvements in precision only on the CLEF collection. 


\par \textbf{RQ3:} \textit{What is the effect of guiding the query formulation process with ChatGPT through multiple prompts that mimic the process of the current state-of-the-art automated Boolean query generation method?}

We have leveraged the process set out by the existing objective query generation method~\cite{scells2020comparison,scells2020computational} to design a sequence of prompts that iteratively guide ChatGPT in the process of generating a Boolean query. Our results show that guided prompts lead to higher effectiveness than single prompt strategies: improvements are observed for both precision and recall. There are caveats though that one need to be wary about -- we discuss these next.


\par \textbf{RQ4:} \textit{What are the caveats and potential challenges of using ChatGPT to create systematic review Boolean queries?} 

Our results highlight two main caveats practitioners should be wary about if relying on ChatGPT to create Boolean queries for systematic review literature search: (1) incorrect MeSH terms, (2) high variability in query effectiveness across multiple requests. 

We already touched upon the first caveat when analysing RQ1. 55\% and over of the MeSH terms generated by ChatGPT are actually not in the MeSH vocabulary, and thus are incorrect. The effect of these incorrect MeSH terms is still unclear, as it is unclear whether existing methods for automatic MeSH term suggestion could be improved to correct the MeSH terms generated by ChatGPT~\cite{wang2021mesh,wang2022automated}.

The second caveat refers to the results showed in Figures~\ref{fig:variability_formulation}-~\ref{fig:variability_guided-s}, which highlight that if multiple generations are performed with ChatGPT for the same prompt, different queries are generated -- and their effectiveness can largely differ. The fact that the output of ChatGPT is non-deterministic given the same prompt\footnote{We validated this by issuing our prompts, as well as through other interactions, across separate ``chats'' to avoid ChatGPT to use the previous input and output as part of the input for the current interaction.} suggests that it may use top-k sampling, or a similar technique, for text generation. Top-k sampling involves generating the $k$ most likely next tokens for a given input and then randomly selecting one of those tokens as the next step in the generation process.
Nevertheless, aside from the specific technique used for text generation, the non-deterministic nature of ChatGPT and the fact that this has a sensible impact of the effectiveness of the Boolean queries that are generated, pose a threat for the uptake of the methods we investigated in this paper. In fact, users would typically have no or limited insight in the effectiveness of the query that ChatGPT generates and thus may be unable to identify high-yield queries among those generated for the same prompt.

\section{Limitations in our Use of ChatGPT}

In our experiment, we use ChatGPT as a black-box application and test its effectiveness in generating systematic review Boolean queries. One significant limitation of our paper is that details about how ChatGPT works internally are undisclosed, including model architecture and what's included in training data. Therefore, we do acknowledge that ChatGPT may have already seen the original queries for the systematic review topics in our test collection during model training and this might impact the effectiveness of the queries ChatGPT generated.
In addition, we are unclear whether interaction history (beyond immediate context of conversation, which the service provider, OpenAI, states being approximately 3000 words or 4000 tokens) may have effect on the generation. Although when executing an experiment for a new query topic we instantiate a new conversation within ChatGPT, we do not know what information the service provider retains for a user and whether they use previous conversations of the user to personalise generation.

Thirdly, our experiments were executed over a two-weeks period ion early January 2023. We are unsure whether modifications to the model have been made during this period. OpenAI disclosed the model we used is the January 9 version and no experiment was run once OpenAI updated their model to the Jan 30 version. Yet, we could not control for minor engineering or model updates OpenAI may have made to the model but not disclosed.

\section{Conclusion}

This paper used ChatGPT to formulate and refine Boolean queries for systematic reviews. We designed an extensive set of prompts to investigate these tasks on over 100 systematic review topics.
Queries generated by ChatGPT obtained higher precision but lowered recall compared to the queries generated by the current state-of-the-art automatic query formulation methods. We also showed that ChatGPT could generate Boolean queries with even higher effectiveness with a guided prompt. 
One caveat of our results is that ChatGPT generates different queries even if the same prompt is used, which vary in effectiveness. This issue would need to be resolved before using ChatGPT to generate Boolean queries for systematic review literature search in practice, where reproducibility is paramount.

In summary, the paper makes the following contributions:

\begin{itemize}[leftmargin=10pt,noitemsep,topsep=0pt]
	\item To the best of our knowledge, this paper is the first of its kind to investigate the effectiveness of Boolean queries designed for systematic review literature search generated by ChatGPT.
	
	\item Several strategies for generating prompts for ChatGP.
	
	\item An extensive evaluation of a large number of systematic review topics and prompts provides insight into how different prompts may impact the effectiveness of the systematic review Boolean queries generated through ChatGPT.
	
	\item A query failure analysis provides possible directions for future research on generating better queries with ChatGPT or post-process results from these queries.
\end{itemize}

Overall, it is still unclear whether all Boolean queries could be developed by transformer-based generative models like ChatGPT. This paper shows the potential for such models, tempered by many issues around variability, robustness and reproducibility. Whatever the outcome, we firmly believe this is an area that is a promising and exciting ground for future research.

\clearpage
\bibliographystyle{ACM-Reference-Format}
\interlinepenalty=10000
\bibliography{sigir-23-generating-queries-chatgpt.bib, harry_thesis_references.bib}

\end{document}